\documentclass[a4paper]{IEEEtran}
\usepackage[utf8]{inputenc}
\usepackage{lipsum}
\usepackage{amsmath,amssymb,amsthm}
\usepackage[linesnumbered,ruled,vlined,titlenumbered]{algorithm2e}
\usepackage{algorithmicx}

\usepackage{cite}

\usepackage{tikz}
\usepackage{pgfplots}

\usepackage{tikzscale}
\pgfplotsset{compat=newest}
\usepackage{xcolor}
\usepackage{mathtools}
\usepackage{url}
\usepackage{booktabs}

\newtheorem{theorem}{Theorem}
\newtheorem{definition}{Definition}

\newtheorem{lemma}{Lemma}

\newtheorem{problem}{Problem}

\def\ve#1{{\mathchoice{\mbox{\boldmath$\displaystyle #1$}}%
		{\mbox{\boldmath$\textstyle #1$}}%
		{\mbox{\boldmath$\scriptstyle #1$}}%
		{\mbox{\boldmath$\scriptscriptstyle #1$}}}}

\IEEEoverridecommandlockouts

\newif\ifcomment

\author{\IEEEauthorblockN{Cornelia Ott\IEEEauthorrefmark{1}, Hedongliang Liu\IEEEauthorrefmark{2}, Antonia Wachter-Zeh\IEEEauthorrefmark{2}\\}
\IEEEauthorblockA{%
\IEEEauthorrefmark{1}Institute of Communications Engineering, Ulm University, Germany \\
\IEEEauthorrefmark{2}Institute for Communications Engineering, Technical University of Munich (TUM), Germany \\
{\small E-mail: cornelia.ott@uni-ulm.de,  lia.liu@tum.de, antonia.wachter-zeh@tum.de}
}
\thanks{The work of H.~Liu and A.~Wachter-Zeh has been supported by a German Israeli Project Cooperation (DIP) grant under grant no.~PE2398/1-1 and KR3517/9-1.}
}
\title{
Covering Properties of Sum-Rank Metric Codes
}

\newcommand{\Fq}{\mathbb{F}_q}
\newcommand{\Fqm}{\mathbb{F}_{q^m}}
\newcommand{\NN}{\mathbb{N}}

\newcommand{\Code}{\mathcal{C}}
\newcommand{\E}{\mathcal{E}_k(\Fqm^n)}
\newcommand{\x}{\ve{x}}
\renewcommand{\c}{\ve{c}}
\newcommand{\G}{\ve{G}}
\newcommand{\I}{\ve{I}}
\newcommand{\X}{\ve{X}}

\newcommand{\A}{\ve{A}}

\newcommand{\y}{\ve{y}}
\renewcommand{\v}{\ve{v}}
\renewcommand{\u}{\ve{u}}
\newcommand{\e}{\ve{e}}
\newcommand{\w}{\ve{w}}
\renewcommand{\b}{\ve{b}}

\newcommand{\wth}{\mathrm{wt}_{SR,n}}
\newcommand{\wtr}{\mathrm{wt}_{SR,1}}

\newcommand{\wtsr}{\mathrm{wt}_{SR,\ell}}
\newcommand{\dsr}{\mathrm{d}_{SR,\ell}}
\newcommand{\dr}{\mathrm{d}_{SR,1}}
\renewcommand{\dh}{\mathrm{d}_{SR,n}}
\renewcommand{\d}{\mathrm{d}}
\newcommand{\Sl}{\mathcal{S}_\ell}
\newcommand{\Bl}{\mathcal{B}_\ell}
\newcommand{\Bone}{\mathcal{B}_{1}}
\newcommand{\Bn}{\mathcal{B}_{n}}

\newcommand{\Ksr}{\mathcal{K}_{SR,\ell}}
\newcommand{\Kr}{\mathcal{K}_{SR,1}}
\newcommand{\Kh}{\mathcal{K}_{SR,n}}
\newcommand{\rhosr}{\rho_{SR,\ell}}
\newcommand{\rhor}{\rho_{SR,1}}
\newcommand{\rhoh}{\rho_{SR,n}}
\newcommand{\VolS}{\mathrm{Vol}_{\mathcal{S}_{\ell}}}
\newcommand{\VolB}{\mathrm{Vol}_{\mathcal{B}_{\ell}}}
\newcommand{\VolI}{\mathrm{Vol}_{\mathcal{I}_{\ell}}}
\newcommand{\rk}{\mathrm{rk}}
\newcommand{\rkq}{\mathrm{rk}_q}

\makeatletter
\newcommand{\removelatexerror}{\let\@latex@error\@gobble}
\makeatother

\IEEEoverridecommandlockouts
\begin{document}

\maketitle
\pagestyle{plain}
\commentfalse
\begin{abstract}
The sum-rank metric can be seen as a generalization of both, the rank and the Hamming metric. It is well known that sum-rank metric codes outperform rank metric codes in terms of the required field size to construct \emph{maximum distance separable} codes (i.e., the codes achieving the Singleton bound in the corresponding metric). In this work, we investigate the covering property of sum-rank metric codes to enrich the theory of sum-rank metric codes. We intend to answer the question: what is the minimum cardinality of a code given a sum-rank covering radius? We show the relations of this quantity between different metrics and provide several lower and upper bounds for sum-rank metric codes.
\end{abstract}

\section{Introduction}
The sum-rank metric was implicitly used in space-time coding schemes~\cite{ElGamalHammons2003space-time,LuKumar2005space-time} and explicitly introduced in network coding literature~\cite{nobrega2010multishot}. It can be seen as a generalization of the Hamming metric and the rank metric.
The theory of sum-rank metric codes attracted a lot of research interest in recent years because of the existence of \emph{maximum sum-rank distance} (MSRD) codes (i.e., codes achieving the Singleton bound in sum-rank metric) with sub-exponential alphabet size in code length, e.g., linearized Reed-Solomon codes \cite{martinez2018skew}. This overcomes the disadvantage of \emph{maximum rank distance} (MRD) codes, which requires exponential alphabet size. Unlike in the rank-metric, non-trivial perfect codes can be constructed in the sum-rank metric~\cite{martinez2020sumrankperfect}.
In \cite{byrne2020fundamental, ott2021bounds} bounds on the cardinality (e.g., sphere-packing and Gilbert-Varshamov bounds) and other properties (e.g., existence and duality of MSRD codes) for sum-rank metric codes have been considered.
Code constructions and decoding algorithm have been extensively studies recently~\cite{martinez2018skew,boucher2019algorithm,martinez2019reliable,caruso2019residues,bartz2020fast,martinezpenas2020sumrank,byrne2020fundamental,puchinger2020generic,Hoermann2022efficient,Hoermann2022errorerasure,Hoermann2022speeding}.
Some of these codes have found applications in distributed data storage \cite{martinez2019universal}, others in aspects of network coding \cite{martinez2019reliable}, and space-time codes \cite{shehadeh2020rate}.
A detailed summary on properties and applications of sum-rank metric codes can be found in a recent survey~\cite{FnTsurvey-Umberto}.

The covering radius is a basic geometric parameter of a code and can be viewed as a measure of performance of a code. To be precise, the covering radius of a code corresponds to the maximum weight of a correctable error by the code.
This quantity has been extensively studied for the Hamming metric (e.g., \cite{cohen1985covering,cohen1997covering,bartoli2014covering}). For the rank metric, the covering property has been studied in \cite{gadouleau2008packing,gadouleau2009bounds}. For the sum-rank metric, this quantity has not been studied so far, to the extent of the knowledge of the authors. In this work we extend the bounds from \cite{gadouleau2008packing} to sum-rank metric codes.

This paper is organized as following: In Section \ref{sec:pre} we introduce the notations used throughout the paper, the sum-rank metric and the covering problem. The relations of covering radius and the minimum cardinality of codes given a covering radius in the rank, the sum-rank and the Hamming metric are shown in Section \ref{sec:covering-sr}. In Section \ref{sec:bounds}, we present the main results of this work, several lower and upper bounds on the minimum cardinality of a codes given a sum-rank covering radius. Finally, we conclude this paper in Section \ref{sec:conclusion}.

\section{Preliminaries}
\label{sec:pre}
Let $q$ be a prime power and $m, n, \ell, \eta$ positive integers, $\Fq$ a finite field with $q$ elements and $\Fqm$ its extension field. We consider in this paper codes as subsets of $\Fqm^n$,
where each vector $\x=[\x_1| \ldots| \x_\ell] \in \Fqm^n$ consists of $\ell$ blocks $\x_1, \ldots, \x_\ell \in \Fqm^\eta$ of length $\eta$. Hence we assume $n=\ell\cdot \eta$. 
Since $\Fqm$ can be seen as an $m$-dimensional vector space over $\Fq$, we can represent a vector $\x_i \in \Fqm^\eta$ as a matrix $\X_i \in \Fq^{m \times \eta}$. The rank of $\x_i$ is defined as the rank of the matrix $\X_i$, i.e., $\rkq(\x_i) \coloneqq \rk(\X_i)$. 
For $\x_i \in \Fqm^\eta$ it holds that $\rkq(\x_i)\in \{0, \ldots, \mu\}$, where $\mu\coloneqq \min\{m,\eta\}$. 

\subsection{Sum-Rank Metric}

\begin{definition}
Let $\x=[\x_1| \ldots| \x_\ell] \in \Fqm^n$. \emph{The ($\ell$-)sum-rank weight} of $\x$ is defined as 
\[
\wtsr: \Fqm^n \rightarrow \NN, \quad \x \mapsto \textstyle\sum_{i=1}^{\ell}\rkq(\x_i).
\]
For two vectors $\x, \x' \in \Fqm^n$ the \emph{($\ell$-)sum-rank distance} is defined as 
\begin{align*}
\dsr: \Fqm^n \times \Fqm^n &\rightarrow  &&\NN, \\ (\x,\x') &\mapsto  &&\dsr(\x,\x')\coloneqq\wtsr(\x-\x').
\end{align*}
For a subspace $\mathcal{V}\subset \Fqm^n$ we define the ($\ell$-)sum-rank distance of a vector $\x \in \Fqm^n\setminus \mathcal{V}$ to the subspace $\mathcal{V}$ as follows:
\begin{align*}
\dsr(\x,\mathcal{V})=\min_{\v \in \mathcal{V} }\{\dsr(\x,\v)\}.
\end{align*}
\end{definition}

The ($\ell$-)sum-rank distance $\dsr$ is a metric over $\Fqm^n$, the so-called \emph{sum-rank metric}. Since ($\ell$-)sum-rank metric becomes rank metric for $\ell=1$ and Hamming metric for $\ell=n$, we denote throughout the paper by $\wtr$, $\wth$, $\dr$ and $\dh$ the weight and the distance in rank metric ($\ell=1$) and in Hamming metric ($\ell=n$), respectively. 

Note that the sum-rank weight of a vector is at most its Hamming weight. This can be seen when considering $\x=[\x_1|\ldots| \x_\ell]\in \Fqm^n$ with $\wth(\x)=n-t=\eta-t_1+\ldots+\eta-t_\ell$ where $\sum_{i=1}^\ell t_i=t$ and each $\x_i$ has $t_i$ zero entries. For the sum-rank weight one gets $\wtsr(\x)=\sum_{i=1}^\ell\rkq(\x_i)\leq \sum_{i=1}^\ell \min\{m,\eta-t_i\}\leq\sum_{i=1}^\ell(\eta-t_i)=n-t=\wth(\x)$.
Moreover the rank weight of a vector is at most its sum-rank weight, since (assuming w.l.o.g. $n\leq m$) a vector $\x\in \Fqm$ of sum-rank weight $t=t_1+\ldots+t_\ell$ (i.e., each $\x_i$ has $t_i$ $\Fq$-linearly independent columns for $i\in\{1,\ldots\,\ell\}$), has at most $t$ $\Fq$-linearly independent columns in the union of all blocks, which corresponds to the rank weight of $\x$. 

In the following we define spheres and balls in the sum-rank metric analogues to \cite{loidreau2006properties} and give definitions for their volume.

\begin{definition}
Let $\tau \in \mathbb{Z}_{\geq 0}$ with $0\leq\tau\leq \ell\cdot \mu$ and $\x \in \Fqm^n$.
The sum-rank metric sphere with radius $\tau$ and center $\x$ is defined as
\[
\Sl(\x,\tau)\coloneqq \{\y\in \Fqm^n \mid \dsr(\x,\y)=\tau\}.
\]
Analogously, we define the ball of sum-rank radius $\tau$ with center $\x$ by
\[
\Bl(\x,\tau) \coloneqq \textstyle\bigcup_{i=0}^{\tau}\Sl(\x,i).
\]
We also define the following cardinalities:
\begin{align*}
\VolS(\tau) &\coloneqq |\{\y\in \Fqm^n \mid \wtsr(\y)=\tau\}|, \\
\VolB(\tau) &\coloneqq \textstyle\sum_{i=0}^{\tau}\VolS(i).
\end{align*}
\end{definition}

The sum-rank metric is invariant under translation of vectors. Hence the volume of a sphere or a ball is independent of its center i.e., $\VolS(\tau)$ and $\VolB(\tau)$ are the volumes of any sphere or ball of radius $\tau$. Moreover we can define the volume of the intersection of two equal sized balls $|\Bl(\x_1,\tau)\cap\Bl(\x_2, \tau)|$ independently of their centers but only dependent on their radii $\tau$ and the distance $\delta \coloneqq\dsr(\x_1,\x_2)$ between their respective centers as follows:
\begin{align*}
\VolI(\tau,\delta)\coloneqq \{\y\in\Fqm^n|\wtsr(\y)\leq \tau \land \dsr(\y,\x)\leq\tau\},
\end{align*}
where $\x\in \Fqm^n$ is arbitrary but with fixed weight $\wtsr(\x)=\delta$. Obviously if $\delta>2\tau$, then $\VolI(\tau,\delta)=0$.

In order to give the exact volume of a sphere of radius $t$ we define the set
\[
\tau_{t,\ell,\mu}\coloneqq\left\{\ve{t}=(t_1, \ldots, t_\ell)\mid \textstyle\sum_{i=1}^\ell t_i =t \land t_i\leq \mu, \forall i\right\},
\]
which corresponds to the set of ordered partitions with bounded number of bounded summands. 
The size of such partitions will be used often throughout the paper, since it corresponds to the number of possibilities how to partition the sum-rank weight $t$ into $\ell$ blocks where each block has at most rank $\mu$.
By the common combinatorial methods, we obtain
\begin{align}
\label{ineq: tau}
|\tau_{t,\ell,\mu}| = \textstyle\sum_{i=0}^\ell (-1)^i \tbinom{\ell}{i}\tbinom{t+\ell-(\mu+1)i}{\ell -1} \leq \tbinom{t+\ell-1}{\ell-1}
\end{align}
(see also \cite[Lemma 1.1]{ratsaby2008estimate}). The upper bound $\binom{t+\ell-1}{\ell-1}$ can also be easily derived by a stars-$\&$-bars argument.

\begin{definition}
A linear $[n,k,d]$ ($\ell$-)sum-rank metric code $\Code$ over $\Fqm$ is an $\Fqm$-vector space $\Code\subset \Fqm^n$ with $\dim_{\Fqm}(\Code)=k$. Hence, the cardinality of the code is $|\Code|=q^{mk}$. Each codeword $\c=[\c_1|\ldots| \c_\ell]\in \Code$ consists of $\ell$ blocks $\c_i\in \Fqm^\eta$ of length $\eta$. The minimum ($\ell$)-sum-rank distance $d$ is defined as
\[
d \coloneqq \min_{\c\neq \c'\in \Code}\{\dsr(\c,\c')\}=\min_{\c \in \Code}\{\wtsr(\c)\}.
\]
\end{definition}

The fact that the sum-rank weight of a vector is at most its Hamming weight implies the following Singleton bound in the sum-rank metric.
\begin{theorem}[\!\!{\cite[Corollary 4, 5]{martinez2019universal}}{\cite[Theorem 5]{martinez2019reliable}}]
\label{Singleton}
Let $\Code$ be a linear $[n,k,d]$ ($\ell$-)sum-rank metric code over $\Fqm$. Then it holds
\[
d\leq \mu\ell-\frac{\mu}{\eta}k+1,
\]
where $\mu=\min\{\eta,m\}$.\end{theorem}
Codes that fulfill this bound with equality are called \emph{maximum sum-rank distance} (MSRD) codes.

\subsection{Covering Property}
In this paper, we study lower and upper bounds on the cardinality of a code given its \emph{covering radius}. 
\begin{definition}
Let $\Code$ be a linear $[n,k,d]$ ($\ell$-)sum-rank metric code over $\Fqm$. The covering radius of $\Code$ is the smallest integer $\rhosr$ such that any vector $\x \in \Fqm^n$ has at most sum-rank distance $\rhosr$ to some codeword $\c\in \Code$,
In other words, $\rhosr$ is the maximal sum-rank distance from any vector to its nearest codeword 
i.e., $\rhosr=\max_{\x\in \Fqm^n}\{\dsr(\x,\Code)\}$. 
Analogously we denote by $\rho=\max_{\x\in \Fqm^n}\{\d(\x,\Code)\}$ the metric independent covering radius. 
\end{definition}
 
For two codes $\Code\subset \Code'\subset \Fqm^n$ it is known that their covering radii $\rho\geq \rho'$, independent of metric. It can be shown that $\rho$ is at least the minimum distance $d'$ of $\Code'$.
To see this, choose $\x\in \Code' \setminus \Code$ which leads to $\d(\x,\c)\geq d', \forall \c\in \Code'$ and therefore $\forall \c \in \Code$. It follows  that $\min_{\c\in \Code}\{\d(\x,\c)\}\geq d'$, i.e., $\d(\x,\Code)\geq d'$. Hence $\rho=\max_{\x\in \Fqm^n}\{\d(\x,\Code)\}\geq d'$.
This leads to the definition of \emph{maximal codes}. 
\begin{definition}
A linear $[n,k,d]$ code $\Code$ is called \emph{maximal} if there is no other $[n,k',d]$ code $\Code'$ such that $\Code\subset \Code'$. 
\end{definition}
It is common knowledge that the covering radius of a maximal code is smaller than its minimum distance. This fact can be seen since for a maximal $[n,k,d]$ code $\Code$ there exists no element $\x\in \Fqm^n\setminus \Code$ such that $\d(\x,\Code)\geq d$; otherwise, one can construct $\Code'=\{\x\}\cup\Code\supset \Code$ with $d(\Code')=d$.
Rather it holds that $\d(\x,\Code)<d, \forall \x \in \Fqm^n\setminus \Code$ and hence  $\rho=\max_{x\in \Fqm^n}\{\d(\x,\Code)\}<d$.

\section{Covering Properties of sum-rank metric codes}
\label{sec:covering-sr}
For a given vectorspace $\Fqm^n$ and a given integer $\rho$ we denote the minimum cardinality of a code $\Code\subset \Fqm^n$ with sum-rank covering radius $\rho$ by $\Ksr(\Fqm^n,\rho)$.
We can express this value as follows:
let $\mathcal{A}\coloneqq \{\Code\subset \Fqm^n: \bigcup_{\c\in\Code} \Bl(\c,\rho)\supset\Fqm^n \}$ then $\Ksr(\Fqm^n,\rho)=\min_{\Code\in\mathcal{A}}\{|\Code|\}$.
We now formulate the sphere covering problem for the sum-rank metric.
\begin{problem}
Find the minimum number of sum-rank balls $\Bl(\x,\rho)$ of radius $\rho$ (with $\x\in \Fqm^n$) that cover the space $\Fqm^n$ entirely. This problem is equivalent to determining the minimum cardinality $\Ksr(\Fqm^n,\rho)$ of a code $\Code\subset \Fqm^n$ with sum-rank covering radius $\rho$. 
\end{problem}
There are two extreme cases for the covering radius: $\Ksr(\Fqm^n,0)=q^{mn}$ and $\Ksr(\Fqm^n,\mu\ell)=1$. The first statement holds, since from $\rhosr=\max_{x\in \Fqm^n}\{\dsr(\x,\Code)\}=0$ it follows that $\dsr(\x,\Code)=0, \forall \x \in \Fqm^n$ and therefore $\x\in \Code$, i.e., $\Code=\Fqm^n$.  
For the second statement we consider $\rhosr=\max_{x\in \Fqm^n}\{\dsr(\x,\Code)\}=\mu\cdot\ell$ which means that there exists an $\x\in \Fqm^n$ such that $\dsr(\x,\Code)=\mu\cdot\ell$. This is already fulfilled by choosing $\Code=\{(0,\ldots,0)\}$. To consider non-trivial cases, we assume $0<\rhosr<\mu\cdot\ell$.

For a fix code $\Code\subset  \Fqm^n$, the relation between the covering radii in rank and Hamming metric is well known. In the following we classify the covering radii in sum-rank metric into this relation.

\begin{lemma} Let $\Code\subset\Fqm^n$ then it holds for its corresponding covering radii $\rhor$, $\rhosr$ and $\rhoh$ in the rank, the sum-rank and the Hamming metric that
$$\rhor\leq\rhosr\leq\rhoh.$$
\end{lemma}
\begin{proof}
Since $\wtr(\x)\leq\wtsr(\x)\leq\wth(\x)$ for a fix $\x\in \Fqm$ it follows that $\dr(\x,\Code)\leq\dsr(\x,\Code)\leq \dh(\x,\Code)$ and hence $\max_{\x\in \Fqm^n}\{\dr(\x,\Code)\}\leq \max_{\x\in \Fqm^n}\{\dsr(\x,\Code)\}\leq \max_{\x\in \Fqm^n}\{\dh(\x,\Code)\}$.
\end{proof}
We give in the following theorem the relation between the minimum cardinality of a code $\Code\subset \Fqm^n$ with a fix covering radius $\rho$ in the rank, the sum-rank metric and the Hamming metric.
\begin{theorem}\label{thm:Krelations}
For $0<\rho< \mu\cdot\ell$, it holds $\Kr(\Fqm^n, \rho)\leq\Ksr(\Fqm^n, \rho)\leq\Kh(\Fqm^n, \rho)$.
\end{theorem}
\begin{proof}
Let $\mathcal{A}_{SR,\ell} \coloneqq \{\Code\subset \Fqm^n| \bigcup_{\c\in\Code} \Bl(\c,\rho)\supset\Fqm^n \}$ 
be the set of codes with sum-rank covering radius $\rho$. 
Since $\wtr(\x)\leq\wtsr(\x)\leq\wth(\x)$ for a fix $\x\in \Fqm$, one gets $\bigcup_{\c\in\Code} \Bone(\c,\rho)\supset \bigcup_{\c\in\Code} \Bl(\c,\rho)\supset \bigcup_{\c\in\Code} \Bn(\c,\rho)$  and hence it follows that $\mathcal{A}_{SR,1}\supset \mathcal{A}_{SR,\ell} \supset \mathcal{A}_{SR,n}$. With $\Ksr(\Fqm^n,\rho)=\min_{\Code\in\mathcal{A}_{SR,\ell}}\{|\Code|\}$ the statement follows.
\end{proof}

\section{Bounds for the Sphere Covering Problem}
\label{sec:bounds}
For the rank metric, a lower and an upper bound on the minimum cardinality of a code with given covering radius was derived in \cite[Proposition 6]{gadouleau2008packing}. The lower bound is called sphere covering bound. We give in the following the sum-rank metric analogue sphere covering bound.
\subsection{Lower Bounds}
\begin{theorem}[Sphere Covering Bound]\label{SPBound}
For the minimum cardinality of a code $\Code\subset \Fqm^n$ with sum-rank covering radius $0<\rho< \mu\cdot\ell$ the following inequality holds: $$\frac{q^{mn}}{\VolB(\rho)}\leq\Ksr(\Fqm^n, \rho).$$ 
\end{theorem}
\begin{proof}
If it is possible to cover the whole space $\Fqm^n$ with balls of radius $\rho$ without overlapping any two balls, then $\frac{q^{mn}}{\VolB(\rho)}=\Ksr(\Fqm^n, \rho)$. This is only possible for perfect sum-rank metric codes. In contrast to the rank metric case where no nontrivial perfect codes exist, in the sum-rank metric such codes indeed exist, since there are nontrivial perfect codes in the Hamming metric, which are included in the sum-rank metric when considering the number of blocks $\ell=n$. If there are overlapping balls then $\frac{q^{mn}}{\VolB(\rho)}<\Ksr(\Fqm^n, \rho)$.
\end{proof}

In order to calculate the bound exactly one needs the exact volume of a ball for a given sum-rank radius. Therefore we need the number of $m\times n$ matrices over $\Fq$ for a given rank $t \leq \min\{m,n\}$, which we denote by
\[
\mathrm{NM}_q(n,m,t) \coloneqq \begin{bmatrix}n\\t\end{bmatrix}_{q} \cdot \prod\textstyle_{i=0}^{t-1}(q^m-q^i)
\]
(see e.g., \cite{migler2004weight}), where $\begin{bmatrix}n\\t\end{bmatrix}_{q}=\prod_{i=1}^{t}\frac{q^{n-t+i}-1}{q^i-1}$ denotes the the Gaussian binomial coefficient, which is is defined by the number of $t$-dimensional subspaces of $\Fq^n$. Hence the volume of a sphere containing all vectors in $\Fqm^n$ of sum-rank weight $t$ is $\VolS(t)=\sum_{\ve{t}\in \tau_{t,\ell,\mu}}\prod_{i=1}^\ell \mathrm{NM}_q(\eta,m,t_i)$ and therefore the volume of a ball of sum-rank radius $t$ is
$$
\VolB(t)=\sum_{j=0}^{t}\sum_{\ve{t}\in \tau_{j,\ell,\mu}}\prod_{i=1}^\ell \mathrm{NM}_q(\eta,m,t_i).
$$
Note that this can be computed in complexity $\tilde{\mathcal{O}}\big(\ell^2t^3+\ell d^t(m+\eta)\log(q)\big)$  using the efficient algorithm for computing $\VolS$ in \cite[Theorem 6 and Algorithm 1]{puchinger2020generic}.

In \cite[Theorem 5]{puchinger2020generic} an upper bound on the sphere size $\VolS(\rho)$ was derived. We use this bound to formulate a simplified version  of the sphere covering bound. Define
\begin{equation*}
\gamma_q := \prod\textstyle_{i=1}^{\infty} (1-q^{-i})^{-1}. 
\end{equation*}
Note that $\gamma_q$ is monotonically decreasing in $q>1$ with a limit of $1$, and e.g., $\gamma_2 \approx 3.463$, $\gamma_3 \approx 1.785$, and $\gamma_4 \approx 1.452$.
\begin{theorem}[Simplified Sphere Covering Bound]
\label{SimplSCBound}
For the minimum cardinality of a code $\Code\subset \Fqm^n$ with sum-rank covering radius $0<\rho< \mu\cdot\ell$ the following inequality holds: $$\frac{q^{mn-\rho(m+\eta-\frac{\rho}{\ell})}}{\rho\cdot\binom{\ell+\rho-1}{\ell-1}\gamma_q^\ell} \leq\Ksr(\Fqm^n, \rho)$$
\end{theorem}
\begin{proof}
In \cite[Theorem 5]{puchinger2020generic} the following upper bound on the sphere size $\VolS(\rho)$ was given: 
\begin{align*}
\VolS(\rho)\leq\binom{\ell+\rho-1}{\ell-1}\gamma_q^\ell q^{\rho(m+\eta-\frac{\rho}{\ell})}.
\end{align*}
Since $\VolB(\rho) = \sum_{\rho'=0}^{\rho} \VolS(\rho') \leq \rho\VolS(\rho)$ for $\rho>1$,  
this gives an upper bound on $\VolB(\rho)$. Together with Theorem~\ref{SPBound} we get the claim by plugging in the upper bound on $\VolB(\rho)$.
\end{proof}

For the rank metric it was shown in \cite[Proposition 7]{gadouleau2009bounds} that a code $\Code\subset \Fqm^n$ with covering radius $0<\rho<n<m$ consists of at least three codewords. We extend this result to the sum-rank metric. In contrast to \cite{gadouleau2009bounds}, there is no restriction on the relation between $n$ and $m$.
\begin{theorem}
For the covering radius $\rho$ fulfilling $0<\rho<\mu\cdot\ell$ the minimum cardinality $\Ksr(\Fqm^n)$ of a code is greater than $3$.
\end{theorem}
\begin{proof}
We assume there exists a  code $\Code\subset\Fqm^n$ of length $n$ with $|\Code|=2$ and covering radius $\rho<\mu\cdot \ell$. W.l.o.g., we suppose $\Code=\{\ve{0},\c\}$. In order to get the smallest possible covering radius $\rho$ we choose $\c$ of sum-rank weight $\mu\ell$ such that the cardinality of the union of the balls with radius $\rho$ around the two codewords $|\Bl(\ve{0},\rho)\cup\Bl(\c,\rho)|$ is maximal. The code $\Code'\coloneqq\langle\c\rangle$ is a linear $[n,1,\mu\ell]$ sum-rank metric code. Therefore any $\x\in \Code'\setminus \Code$ has sum-rank distance $\mu\ell$ to $\Code$, i.e., $\rho=\mu\ell$, which is a contradiction to the assumption $\rho<\mu\cdot\ell$. Note that choosing the sum-rank weight of $\c$ smaller cannot lead to a smaller covering radius.
\end{proof}

In the following we give a nontrivial lower bound on the minimum cardinality of a code with given covering radius, which is the sum-rank metric analogue bound to the bound derived in \cite[Proposition 8]{gadouleau2008packing}.
A special feature of this bound is, that the right hand side is a  monotonically increasing function in $k$, where  $0<k\leq\lfloor \log_{q^m}(\Ksr(\Fqm^n,\rho))\rfloor$. This fact allows to use the bound iteratively. We start with $k=\lfloor \log_{q^m}(\Ksr(\Fqm^n,\rho))\rfloor$, for which the bound is the tightest. Since we don't know $\Ksr(\Fqm^n,\rho)$ in general, we use an already known lower bound to calculate the start value of $k$, for instance the simplified sphere covering bound in Theorem \ref{SimplSCBound}. In the $i$-th iteration we use the bound derived in iteration $i-1$ to calculate the new value of $k$. We stop when the value of $k$ cannot be improved anymore by applying the bound.

\begin{theorem}\label{thm:lowerbound}
Let $0<\rho<\mu\cdot \ell$ and $0<k\leq\lfloor \log_{q^m}(\Ksr(\Fqm^n,\rho))\rfloor$ then
\begin{align*}
&\Ksr(\Fqm^n,\rho)\geq  \frac{1}{\VolB(\rho)-\VolI(\rho, \mu\ell-\frac{\mu}{\eta}k)}\\
&\cdot\Big(q^{mn}-q^{km}\VolI(\rho, \mu\ell-\frac{\mu}{\eta}k)+\VolI(\rho,\mu\ell-\frac{\mu}{\eta}k'+1)\\
&\cdot\textstyle{\sum}_{k'=\max\{1,n-2\frac{\eta}{\mu}\rho+1\}}^{k}(q^{k'm}-q^{(k'-1)m})\Big).  
\end{align*}
\end{theorem}
\begin{proof}
Let $k_{\max}\coloneqq\lfloor \log_{q^m}(\Ksr(\Fqm^n,\rho))\rfloor$, $K=q^{mk_{\max}}$ and $\Code\coloneqq\{\c_0,\ldots, \c_{K-1}\}\subset\Fqm^n$ a code with covering radius $\rho$ and $\Code_j\coloneqq\{\c_0,\ldots, \c_{j}\}\subset\Code$ with $0\leq j  \leq K-1$ such that $\dsr(\c_j,\Code_{j-1})\geq\dsr(\c_{j+1},\Code_{j})$ for $\forall j>0$. 
Therefore $\dsr(\c_j,\Code_{j-1})=\min_{1\leq i\leq j}\{ \dsr(\c_i,\Code_{i-1})\}$ is the minimum distance $d_{min}(\Code_j)$ of $\Code_j$. For $1\leq k' \leq k_{\max}$ and $q^{m(k'-1)}\leq j \leq q^{m k'}$, the Singleton Bound for the sum-rank metric (c.f. Theorem \ref{Singleton}) provides $\d_{\min}(\Code_j)=\dsr(\c_j,\Code_{j-1})\leq \mu\cdot\ell-\frac{\mu}{\eta}k'+1$. 
Hence the number of vectors covered by a ball of radius $\rho$ around $\c_j$ that are not covered by the union of balls of radius $\rho$ around the codewords in  $\Code_{j-1}$ is at most $\VolB(\rho)-\VolI(\rho, \mu\cdot\ell-\frac{\mu}{\eta}k'+1)$. 
Since the covering radius of $\Code$ is $\rho$, the number of vectors covered by the union of balls of radius $\rho$ around the codewords in  $\Code$ is $q^{mn}$ and therefore it follows, that
\begin{align*}
    q^{mn}\leq  &\VolB(\rho)+\sum_{k'=1}^{k}(q^{k'm}-q^{(k'-1)m})\\
    &\cdot\Big(\VolB(\rho)-\VolI(\rho, \mu\ell-\frac{\mu}{\eta}k'+1)\Big)\\
    &+(K-q^{km})\Big(\VolB(\rho)-\VolI(\rho, \mu\ell-\frac{\mu}{\eta}k)\Big).
\end{align*}

We transform the inequality to
\begin{align*}
    K\Bigg(&\VolB(\rho)-\VolI(\rho, \mu\ell-\frac{\mu}{\eta}k)\Bigg)\geq\\
    & q^{mn}-\VolI\Big(\rho, \mu\ell-\frac{\mu}{\eta}k\Big)q^{km}\\
    &+\sum_{k'=1}^{k}\Big(q^{k'm}-q^{(k'-1)m}\Big)\Bigg(\VolI\Big(\rho, \mu\ell-\frac{\mu}{\eta}k'+1\Big)\Bigg)\\
    &-\VolB(\rho)\Bigg(\sum_{k'=1}^{k}\Big(q^{k'm}-q^{(k'-1)m}\Big)-q^{km}+1\Bigg).
\end{align*}
Since $\VolI(\rho, \mu\ell-\frac{\mu}{\eta}k'+1)=0$ for $\frac{\mu}{\eta}k'<\mu\ell-2\rho$ this leads to
\begin{align*}
    K\Big(&\VolB(\rho)-\VolI\Big(\rho, \mu\ell-\frac{\mu}{\eta}k\Big)\Big)\geq \\
    &q^{mn}-\VolI\Big(\rho, \mu\ell-\frac{\mu}{\eta}k\Big)q^{km}\\
    &+\sum^{k}_{\substack{k'=\max\{1,\\n-2\frac{\eta}{\mu}\rho+1\}}}\Big(q^{k'm}-q^{(k'-1)m}\Big)
    \cdot\VolI\Big(\rho, \mu\ell-\frac{\mu}{\eta}k'+1\Big)\\
    &-\VolB(\rho)\Big(q^{km}-1-q^{km}+1\Big).
\end{align*}
This proofs the claim.
\end{proof}
To derive an explicit lower bound from Theorem~\ref{thm:lowerbound} we need an expression for the intersections of two balls $\VolI\Big(\rho, \mu\ell-\frac{\mu}{\eta}k\Big)$ in the sum-rank metric. Deriving such an expression is an ongoing work.

\subsection{Upper Bounds}

An upper bound on the minimum cardinality of a code with given covering radius for the rank metric case was derived in \cite[Proposition 6]{gadouleau2008packing}. In the following we extend the bound to the sum-rank metric. 
\begin{theorem}
For the minimum cardinality of a code $\Code\subset \Fqm^n$ with sum-rank covering radius $0<\rho< \mu\cdot\ell$ the following inequality holds: $\Ksr(\Fqm^n, \rho)\leq q^{m(n-\rho)}$. 
\end{theorem}
\begin{proof}
Consider a systematic generator matrix $\G=(\I|\A)$ of a code $\Code$. For each vector $\x=(x_1,\ldots, x_n)\in \Fqm^n$ there exists a codeword $\c=(x_1,\ldots, x_{k},c_{k+1}, \ldots, c_n)\in \Code$ with $\dsr(\x,\c)=\wtsr(0,\ldots, 0,c_{k+1},\ldots, c_n)\leq\wth(0,\ldots, 0,c_{k+1},\ldots, c_n)\leq n-k$. Therefore $\min_{c\in \Code}\{\dsr(\x,\c)\}\leq n-k$ for each $\x \in\Fqm^n$ and hence $\rho=\max_{\x\in \Fqm^n}\{\dsr(\x,\Code)\}\leq n-k$. This leads to the upper bound $\Ksr(\Fqm^n, \rho)\leq|\Code|=q^{mk}\leq q^{m(n-\rho)}$. 
\end{proof}

To prove the next results we require the following definition of \emph{Elementary Linear subspaces} and their properties.
\begin{definition}
Let $\mathcal{V}\coloneqq\langle \b_1, \ldots, \b_k\rangle \subset \Fqm^n$ with $\dim(\mathcal{V})=k$ and $\b_i\in \Fq^n,\forall i \in \{1, \ldots, k\}$ then $\mathcal{V}$ is called an \emph{Elementary Linear subspace} of $\Fqm^n$.
We denote the set of all Elementary Linear subspaces of  of $\Fqm^n$ with $\dim(\mathcal{V})=k$ by $\E$.
\end{definition}

\begin{lemma}\cite{gadouleau2009bounds}
\label{lemma ELS}
Let $\v \in \Fqm^n$ and let $\mathcal{E}$ be the set of all Elementary Linear subspaces of $\Fqm^n$ dimension of $k$. Then $\wtr(\v)\leq k$ if and only if $\v \in  \mathcal{E}$.
\end{lemma}

\begin{lemma}\cite[Lemma 1]{gadouleau2009bounds}\label{unique ELS}
Each vector $\x\in \Fqm^n$ with $\wtr(\x)=k$ belongs to a unique Elementary Linear subspace $\mathcal{V}\in \E$.
\end{lemma}

\begin{lemma}\cite[Lemma 10]{gadouleau2008packing}\label{bijective function}
Let $u>1$, $r<n$, $\mathcal{V}\in \mathcal{E}_r(\Fqm^n)$ and let 
$
f_u: \Fqm^n\rightarrow \mathbb{F}_{q^{m+u}}^n, \quad (v_0, \ldots, v_{n-1})\mapsto (f_u(v_0), \ldots, f_u(v_{n-1}))
$ 
be a linear bijecive mapping such that $f(\Fq)=\Fq$. Then the image $f(\mathcal{V})$ is a subset of an Elementary Linear subspace of $\mathbb{F}_{q^{m+u}}^n$ of dimension $r$. Moreover f is a bijection between $\mathcal{E}_r(\Fqm^n)$ and  $\mathcal{E}_r(\mathbb{F}_{q^{m+u}}^n)$.
\end{lemma}

Lemma \ref{lemma ELS} draws the connection between Elementary Linear subspaces of $\Fqm^n$ and the rank weight of a vector in $\Fqm^n$. We extend this result in Lemma~\ref{lem:ELS-sr} to the relation between a sequence of Elementary Linear subspaces  of $\Fqm^\eta$ and the sum-rank weight of a vector in $\Fqm^n$, where each block is in $\Fqm^\eta$. 
\begin{lemma}\label{lem:ELS-sr}
Let $\v=(\v_1|\ldots|\v_\ell) \in\Fqm^n$ with $\v_i \in \Fqm^\eta, \forall i\in\{1, \ldots,\ell\}$  then $\wtsr(\v)\leq k$ if and only if there are Elementary Linear subspaces $\mathcal{V}_1, \ldots, \mathcal{V}_\ell$ of $\Fqm^\eta$ with $\v_i\in\mathcal{V}_i, \forall i\in\{1, \ldots,\ell\}$ such that $\sum_{i=1}^{\ell}\dim(\mathcal{V}_i)=k$.
\end{lemma}
\begin{proof}
Follows directly by Lemma \ref{lemma ELS}.
\end{proof}
The following Lemma gives a necessary and sufficient condition for a code to be MSRD in terms of Elementary Linear subspaces. A similar result for the rank metric was derived in \cite[Lemma 9]{gadouleau2008packing}. Instead of considering a single Elementary Linear subspace of $\Fqm^n$ we consider for the sum-rank metric the cartesian product of $\ell$ Elementary Linear subspaces of $\Fqm^\eta$. In contrast to \cite{gadouleau2008packing} we have no restriction on the relation between $n$ and $m$.
\begin{lemma}\label{lem:MSRD_ELS}
Let $\Code \subset \Fqm^n$ be an $\ell$-sum-rank metric code of dimension $k$. $\Code$ is an MSRD code if and only if $ \Big(\mathcal{V}_1\times \ldots\times \mathcal{V}_\ell\Big)\oplus \Code=\Fqm^n$ for each sequence $(\mathcal{V}_i)_{1\leq i \leq  \ell}$ fulfilling that each $\mathcal{V}_i$ is an Elementary Linear subspace of $\Fqm^\eta$ and $\sum_{i=1}^{\ell}\dim(\mathcal{V}_i)=\mu\ell-\frac{\mu}{\eta}k$.
\end{lemma}
\begin{proof}
  Let $\Code$ be an MSRD code then following Theorem \ref{Singleton} it holds that $d= \mu\ell-\frac{\mu}{\eta}k+1$. Let $(\mathcal{V}_i)_{1\leq i \leq  \ell}$ be a sequence of Elementary Linear subspaces of $\Fqm^\eta$ fulfilling that
$\sum_{i=1}^{\ell}\dim(\mathcal{V}_i)=\mu\ell-\frac{\mu}{\eta}k$ then 
each $v\in \mathcal{V}_1\times \ldots\times \mathcal{V}_\ell$ has sum-rank weight at most $\mu\ell-\frac{\mu}{\eta}k$ and therefore $\mathcal{V}_1\times \ldots\times \mathcal{V}_\ell\cap \Code=\{0\}$, i.e., $\mathcal{V}_1\times \ldots\times \mathcal{V}_\ell\oplus \Code=\Fqm^n$. Conversely let $(\mathcal{V}_i)_{1\leq i \leq  \ell}$ be a sequence of Elementary Linear subspaces of $\Fqm^\eta$ with
 $\sum_{i=1}^{\ell}\dim(\mathcal{V}_i)=\mu\ell-\frac{\mu}{\eta}k$ and let $\Big(\mathcal{V}_1\times \ldots\times \mathcal{V}_\ell\Big)\oplus \Code=\Fqm^n$. Then $\Big(\mathcal{V}_1\times \ldots\times \mathcal{V}_\ell\Big)\cap \Code=\{0\}$, i.e., $\Code$ contains only vectors of sum-rank weight at least $\mu\ell-\frac{\mu}{\eta}k+1$.
\end{proof}
In order to derive a non-trivial upper bound we need the following result which enables us to construct a code with a given covering radius using an MSRD code over a smaller field with the same covering radius, length and dimension. This construction was proposed for the rank metric in \cite[Proposition 10]{gadouleau2008packing}. In contrast to \cite{gadouleau2008packing} we do not assume that $m\geq n$ but we need a more stringent condition on $\rho$.

\begin{theorem}\label{constructMSRDCode}
Let $\Code\subset \Fqm^n$ be an $[n,n-\rho,\rho+1]$ MSRD code having covering radius $\rho$ and $f$ be a function defined as in Lemma~\ref{bijective function} and $0\leq u\cdot \ell \leq \rho$. Then $f(\Code)$ is an $[n,(n-\rho)]$ $\ell$-sum rank metric code over $\mathbb{F}_{q^{m+u}}$ with covering radius $\rho$.
\end{theorem}
\begin{proof}
Since $f_u: \Fqm^n\rightarrow \mathbb{F}_{q^{m+u}}^n$ is a linear bijecive mapping $f(\Code)\subset \mathbb{F}_{q^{m+u}}^n$ is a linear code of length $n$ and dimension $n-\rho$. 
We show now that the covering radius of $f(\Code)$ is $\rho$. 
Let $\mathcal{A}
$  and $\mathcal{B}
$ be subspaces of $\mathbb{F}_{q}^{m+u}$ over $\Fq$ with $\dim(\mathcal{A})=m
$ and $\dim(\mathcal{B})=u
$, where 
$\mathcal{A} \oplus \mathcal{B}
=\mathbb{F}_{q}^{m+u}$.
We can now express every $\u=[\u_1|\ldots|\u_\ell]\in \mathbb{F}_{q^{m+u}}^n$ as  $\u=\v+\w=[\v_1+\w_1|\ldots|\v_\ell+\w_\ell]$ with $\v_i\in \mathcal{A}^\eta$, 
$\v\in \mathcal{A}^n$, $\w_i\in\mathcal{B}^\eta$ 
and $\w\in\mathcal{B}$. So $\wtsr(\w)=\sum_{i=1}^\ell \wtr(\w_i)\leq \ell\cdot
u$.
Lemma \ref{unique ELS} implies that any $\w_i$ belongs to a unique Elementary Linear subspace $\mathcal{W}_i\in \mathcal{E}_{u}(\mathbb{F}_{q^{m+u}}^\eta)$ for all $i \in \{1,\ldots,\ell\}$. Moreover the  cartesian product of these Elementary Linear subspaces $\mathcal{W}\coloneqq\mathcal{W}_1\times\ldots\times\mathcal{W}_\ell$ is unique as well and $\mathcal{W}\in\mathcal{E}_\rho(\mathbb{F}_{q^{m+u}}^n)$.
Since $\Code$ is MSRD Lemma \ref{lem:MSRD_ELS} implies for any $\mathcal{V}\in \mathcal{M}\coloneqq\{ \mathcal{V}_1\times \ldots\times \mathcal{V}_\ell | \mathcal{V}_i \in \mathcal{E}_\eta(\Fqm^\eta) \land \sum_{i=1}^{\ell}\dim(\mathcal{V}_i)=\mu\ell-\frac{\mu}{\eta}k\}$ that $\mathcal{V}\oplus\Code=\Fqm^n$. 
Since $\mu\ell-\frac{\mu}{\eta}\leq\rho$ it holds that $\mathcal{M}\subset \mathcal{E}_\rho(\Fqm^n)$.
Together with Lemma \ref{bijective function} it follows that we can express $\v=f(\c+\e)=f(\c)+f(\e)$ with $\c \in \Code$ and $\e=[\e_1|\ldots|\e_\ell]$ with $\e\in \mathcal{V}$ with $\mathcal{V}\in\mathcal{M}$ and therefore $f(\mathcal{V})\subset\mathcal{W}$. 
Hence $f(\e)\in \mathcal{W}$ and so $f(\e)+\w\in \mathcal{W}$. So one get $\u=\v+\w=f(\c)+f(\e)+\w$ with $\wtsr(\w+f(\e))=\sum_{i=1}^{\ell}\wtr(\w_i+f(\e_i))\leq \rho$.
Hence
$\dsr(\u,f(\c))\leq \rho$ and therefore $\dsr(\u,f(\Code))\leq \rho$, so $\max_{\u\in\mathbb{F}_{q^m+u}^n}\dsr(\u,f(\Code))\leq \rho$. This shows that the covering radius of $f(\Code)$ is smaller than or equal to $\rho$.
Since $\Code$ has covering radius $\rho$, there is one $\x\in\Fqm^n$ such that $\dsr(\x,\Code)=\rho$. Using that $f$ is rank preserving and since $f$ is also linear $f$ is even sum-rank preserving.Therefore one gets for this $\x$ that $\rho=\dsr(\x,\Code)=\min_{ \c\in\Code}\{\wtsr(\c-\x)\}=\min_{ \c\in\Code}\{\wtsr(f(\c-\x))\}=\min_{ \c\in\Code}\{\wtsr(f(\c)-f(\x))\}=\dsr(f(\x),f(\Code))$, i.e., there is a vector having sum-rank distance $\rho$ to $f(\Code)$. So the covering radius of $f(\Code)$ is exactly $\rho$.
\end{proof}
Finally we give a non-trivial upper bound on the minimum cardinality of a code analogue to the result \cite[Corollary 6]{gadouleau2008packing}. Without any restriction on the relation between $m$ and $n$. 
\begin{theorem}\label{NonTrivialUpperBound}
Let $0\leq\rho\leq \mu\cdot\ell$ then 
$$
\Ksr(\Fqm^{n},\rho)\leq q^{(m-\lfloor\frac{\rho}{\ell}\rfloor)\cdot(n-\rho)}.
$$
\end{theorem}
\begin{proof}
We can construct a linear MSRD code $\Code$ over $\mathbb{F}_{q^{\nu}}^n$ of length $n$, dimension $n-k$ and covering radius $\rho$, 
where $\nu\coloneqq m-\lfloor\frac{\rho}{\ell}\rfloor$. 
Let $f: \mathbb{F}_{q^\nu}^n\rightarrow\Fqm^n$ be a sum-rank preserving mapping. Using Theorem \ref{constructMSRDCode}, $f(\Code)$ has covering radius $\rho$.
So one gets
$$
\Ksr(\Fqm^n, \rho)\leq |f(\Code)|=|\Code|=q^{(m-\lfloor\frac{\rho}{\ell}\rfloor)\cdot(n-\rho)}.
$$
\end{proof}

In the following we apply \cite[Lemma 7]{gadouleau2009bounds} to the sum-rank metric. 
\begin{lemma}
\label{lemmaproduct}
Let $n$, $n'$, $\rho$, $\rho'$, $m$ be  nonnegative integers and $m>0$, then
$\Ksr(\Fqm^{n+n'},\rho+\rho')
\leq \Ksr(\Fqm^n,\rho)
\cdot \Ksr(\Fqm^{ n'},\rho')$. 
\end{lemma}
\begin{proof}
Let $\x, \y \in \Fqm^n$ and $\x', \y' \in \Fqm^{n'}$. We consider $(\x|\x'), (\y|\y') \in \Fqm^{n+n'}$. Then it holds that $\dsr((\x|\x'), (\y|\y'))\leq \dsr(\x,\y)+\dsr(\x',\y')$. Note that equality holds if $\frac{n+n'}{\ell}$ divides $n$ (and therefore as well $n'$). For any linear subspaces $\Code\subset \Fqm^n$ and $\Code'\subset \Fqm^{n'}$ it holds for the sum-rank covering radii of the two subspaces $\Code, \Code'$ and their direct sum $\Code\oplus\Code'$ that $ \rhosr(\Code\oplus\Code')\leq\rhosr(\Code)+\rhosr(\Code')$. 
\end{proof}
Using Lemma \ref{lemmaproduct} we can give now one more upper bound analogue to the bound for the rank metric case given in \cite[Proposition 11]{gadouleau2008packing}. For the sum-rank metric the bound is dependent on the number of blocks $\ell$.

\begin{theorem}
Let $m$, $n$, $\rho$ be fixed positive integers, then for any $l$ with $0\leq l\leq n$ and for every pair $(n_i, \rho_i)$ fulfilling the following three conditions
\begin{itemize}
    \item [(i)]$0 < n_i\leq n$
    \item[(ii)]$0\leq \rho_i \leq n_i$
    \item[(iii)]$n_i+\rho_i\leq m$
\end{itemize}
for all $0\leq i \leq l-1$ with $\sum_{i=0}^{l-1}n_i=n$ and $\sum_{i=0}^{l-1}\rho_i=\rho$ it holds
\begin{align*}
\Ksr(\Fqm^{n},\rho)\leq\min_{l\in \{0,\ldots,n\}}q^{m(n-\rho)-\sum_{i=0}^{l-1}(\lfloor\frac{\rho_i}{\ell}\rfloor)\cdot(n_i-\rho_i)}
\end{align*}

\end{theorem}
\begin{proof}
Using Lemma \ref{lemmaproduct} we have
$\Ksr(\Fqm^{n},\rho)=\Ksr(\Fqm^ {\sum_{i=0}^{l-1}n_i},\sum_{i=0}^{l-1}\rho_i)\leq\prod_{i=0}^{l-1}\Ksr(\Fqm^{n_i},\rho_i)$ for all sequences $(n_i, \rho_i)_{0\leq i \leq l-1}$ with $0\leq l\leq n$ fulfilling the conditions (i)-(iii).
With Theorem \ref{NonTrivialUpperBound} one get for all $i\in\{0,\ldots,l-1\}$ with $0\leq l\leq n$ that $\Ksr(\Fqm^{n_i},\rho_i)\leq q^{(m-\lfloor\frac{\rho_i}{\ell}\rfloor)\cdot(n_i-\rho_i)}$ and therefore
$\Ksr(\Fqm^{n},\rho)
\leq \prod_{i=0}^{l-1} q^{(m-\lfloor\frac{\rho_i}{\ell}\rfloor)\cdot(n_i-\rho_i)}
=q^{\sum_{i=0}^{l-1}(m-\lfloor\frac{\rho_i}{\ell}\rfloor)\cdot(n_i-\rho_i)}
=q^{m(n-\rho)-\sum_{i=0}^{l-1}(\lfloor\frac{\rho_i}{\ell}\rfloor)\cdot(n_i-\rho_i)}
\leq \min_{l\in \{0,\ldots,n\}}q^{m(n-\rho)-\sum_{i=0}^{l-1}(\lfloor\frac{\rho_i}{\ell}\rfloor)\cdot(n_i-\rho_i)}$.
\end{proof}
\section{Conclusions and Future Work}
\label{sec:conclusion}
The sum-rank metric can be seen as a metric that interpolates between the rank and the Hamming metric. In this work, we studied the covering properties of sum-rank metric, in terms of answering the question: what is the minimum cardinality $\Ksr(\Fqm^{n},\rho)$ of a code given a sum-rank covering radius. The relations of this quantity in the rank, the sum-rank and the Hamming metric (see Theorem \ref{thm:Krelations}) provide trivial bounds for $\Ksr(\Fqm^{n},\rho)$ in sum-rank by the known results in the rank and the Hamming metric. In \cite{gadouleau2008packing} non-trivial bounds on the covering property in rank metric have been given. We extended these bounds to sum-rank metric. The advantages are that the field extension degree does not have to be more than the code length in sum-rank metric and we have more flexibility to choose different block size and block length of a code to obtain the best bound. Moreover, in the lower bound on $\Ksr(\Fqm^{n},\rho)$ in Theorem \ref{SimplSCBound}, we used an upper bound on the ball size to derive a simplified lower bound which eases the computation.

An open problem left for the lower bound in Theorem \ref{thm:lowerbound} is the size of the intersection of two balls, which is ongoing work.
In the future, the bounds can be numerically compared for various parameters and to see the gap between the newly derived upper and lower bounds.

\bibliographystyle{IEEEtran}
\bibliography{refs}

\begin{thebibliography}{10}
\providecommand{\url}[1]{#1}
\csname url@samestyle\endcsname
\providecommand{\newblock}{\relax}
\providecommand{\bibinfo}[2]{#2}
\providecommand{\BIBentrySTDinterwordspacing}{\spaceskip=0pt\relax}
\providecommand{\BIBentryALTinterwordstretchfactor}{4}
\providecommand{\BIBentryALTinterwordspacing}{\spaceskip=\fontdimen2\font plus
\BIBentryALTinterwordstretchfactor\fontdimen3\font minus
  \fontdimen4\font\relax}
\providecommand{\BIBforeignlanguage}[2]{{%
\expandafter\ifx\csname l@#1\endcsname\relax
\typeout{** WARNING: IEEEtran.bst: No hyphenation pattern has been}%
\typeout{** loaded for the language `#1'. Using the pattern for}%
\typeout{** the default language instead.}%
\else
\language=\csname l@#1\endcsname
\fi
#2}}
\providecommand{\BIBdecl}{\relax}
\BIBdecl

\bibitem{ElGamalHammons2003space-time}
H.~El~Gamal and A.~Hammons, ``On the design of algebraic space-time codes for
  mimo block-fading channels,'' \emph{IEEE Transactions on Information Theory},
  vol.~49, no.~1, pp. 151--163, 2003.

\bibitem{LuKumar2005space-time}
H.-F. Lu and P.~Kumar, ``A unified construction of space-time codes with
  optimal rate-diversity tradeoff,'' \emph{IEEE Transactions on Information
  Theory}, vol.~51, no.~5, pp. 1709--1730, 2005.

\bibitem{nobrega2010multishot}
R.~W. N{\'o}brega and B.~F. Uch{\^o}a-Filho, ``Multishot codes for network
  coding using rank-metric codes,'' in \emph{2010 Third IEEE International
  Workshop on Wireless Network Coding}.\hskip 1em plus 0.5em minus 0.4em\relax
  IEEE, 2010, pp. 1--6.

\bibitem{martinez2018skew}
U.~Mart{\'\i}nez-Pe{\~n}as, ``Skew and linearized reed--solomon codes and
  maximum sum rank distance codes over any division ring,'' \emph{Journal of
  Algebra}, vol. 504, pp. 587--612, 2018.

\bibitem{martinez2020sumrankperfect}
\BIBentryALTinterwordspacing
U.~Mart\'{\i}nez-Pe\~{n}as, ``Hamming and simplex codes for the sum-rank
  metric,'' \emph{Des. Codes Cryptography}, vol.~88, no.~8, p. 1521–1539, aug
  2020. [Online]. Available:
  \url{https://doi-org.eaccess.ub.tum.de/10.1007/s10623-020-00772-5}
\BIBentrySTDinterwordspacing

\bibitem{byrne2020fundamental}
E.~Byrne, H.~Gluesing-Luerssen, and A.~Ravagnani, ``Fundamental properties of
  sum-rank-metric codes,'' \emph{IEEE Transactions on Information Theory},
  vol.~67, no.~10, pp. 6456--6475, 2021.

\bibitem{ott2021bounds}
C.~Ott, S.~Puchinger, and M.~Bossert, ``Bounds and genericity of
  sum-rank-metric codes,'' in \emph{2021 XVII International Symposium" Problems
  of Redundancy in Information and Control Systems"(REDUNDANCY)}.\hskip 1em
  plus 0.5em minus 0.4em\relax IEEE, 2021, pp. 119--124.

\bibitem{boucher2019algorithm}
D.~Boucher, ``An algorithm for decoding skew reed-solomon codes with respect to
  the skew metric,'' in \emph{Workshop on Coding and Cryptography}, 2019.

\bibitem{martinez2019reliable}
U.~Mart{\'\i}nez-Pe{\~n}as and F.~R. Kschischang, ``Reliable and secure
  multishot network coding using linearized reed-solomon codes,'' \emph{IEEE
  Transactions on Information Theory}, 2019.

\bibitem{caruso2019residues}
X.~Caruso, ``{Residues of Skew Rational Functions and Linearized Goppa
  Codes},'' \emph{arXiv preprint arXiv:1908.08430}, 2019.

\bibitem{bartz2020fast}
H.~Bartz, T.~Jerkovits, S.~Puchinger, and J.~Rosenkilde, ``{Fast Decoding of
  Codes in the Rank, Subspace, and Sum-Rank Metric},'' \emph{arXiv preprint
  arXiv:2005.09916}, 2020.

\bibitem{martinezpenas2020sumrank}
U.~Martínez-Peñas, ``Sum-rank bch codes and cyclic-skew-cyclic codes,'' 2020.

\bibitem{puchinger2020generic}
S.~Puchinger, J.~Renner, and J.~Rosenkilde, ``Generic decoding in the sum-rank
  metric,'' \emph{IEEE Transactions on Information Theory}, pp. 1--1, 2022.

\bibitem{Hoermann2022efficient}
\BIBentryALTinterwordspacing
F.~H{\"o}rmann and H.~Bartz, ``Efficient decoding of folded linearized
  reed-solomon codes in the sum-rank metric,'' in \emph{WCC 2022: The Twelfth
  International Workshop on Coding and Cryptography}, March 2022. [Online].
  Available: \url{https://elib.dlr.de/146410/}
\BIBentrySTDinterwordspacing

\bibitem{Hoermann2022errorerasure}
\BIBentryALTinterwordspacing
F.~Hörmann, H.~Bartz, and S.~Puchinger, ``Error-erasure decoding of linearized
  reed-solomon codes in the sum-rank metric,'' 2022. [Online]. Available:
  \url{https://arxiv.org/abs/2202.06758}
\BIBentrySTDinterwordspacing

\bibitem{Hoermann2022speeding}
\BIBentryALTinterwordspacing
F.~H{\"o}rmann, H.~Bartz, and S.~Puchinger, ``Speeding up error-erasure
  decoding of linearized reed-solomon codes in the sum-rank metric,'' in
  \emph{Coding theory and cryptography: A conference in honor of Joachim
  Rosenthal's 60th birthday}, July 2022. [Online]. Available:
  \url{https://elib.dlr.de/187202/}
\BIBentrySTDinterwordspacing

\bibitem{martinez2019universal}
U.~Mart{\'\i}nez-Pe{\~n}as and F.~R. Kschischang, ``Universal and dynamic
  locally repairable codes with maximal recoverability via sum-rank codes,''
  \emph{IEEE Transactions on Information Theory}, 2019.

\bibitem{shehadeh2020rate}
M.~Shehadeh and F.~R. Kschischang, ``{Rate-Diversity Optimal Multiblock
  Space-Time Codes via Sum-Rank Codes},'' in \emph{IEEE International Symposium
  on Information Theory (ISIT)}, 2020.

\bibitem{FnTsurvey-Umberto}
\BIBentryALTinterwordspacing
U.~Martínez-Peñas, M.~Shehadeh, and F.~R. Kschischang, ``Codes in the
  sum-rank metric: Fundamentals and applications,'' \emph{Foundations and
  Trends® in Communications and Information Theory}, vol.~19, no.~5, pp.
  814--1031, 2022. [Online]. Available:
  \url{http://dx.doi.org/10.1561/0100000120}
\BIBentrySTDinterwordspacing

\bibitem{cohen1985covering}
G.~Cohen, M.~Karpovsky, H.~Mattson, and J.~Schatz, ``Covering radius---survey
  and recent results,'' \emph{IEEE Transactions on Information Theory},
  vol.~31, no.~3, pp. 328--343, 1985.

\bibitem{cohen1997covering}
G.~Cohen, I.~Honkala, S.~Litsyn, and A.~Lobstein, \emph{Covering codes}.\hskip
  1em plus 0.5em minus 0.4em\relax Elsevier, 1997.

\bibitem{bartoli2014covering}
D.~Bartoli, M.~Giulietti, and I.~Platoni, ``On the covering radius of mds
  codes,'' \emph{IEEE Transactions on Information Theory}, vol.~61, no.~2, pp.
  801--811, 2014.

\bibitem{gadouleau2008packing}
M.~Gadouleau and Z.~Yan, ``Packing and covering properties of rank metric
  codes,'' \emph{IEEE Transactions on Information Theory}, vol.~54, no.~9, pp.
  3873--3883, 2008.

\bibitem{gadouleau2009bounds}
------, ``Bounds on covering codes with the rank metric,'' \emph{IEEE
  Communications Letters}, vol.~13, no.~9, pp. 691--693, 2009.

\bibitem{loidreau2006properties}
P.~Loidreau, ``Properties of codes in rank metric,'' in \emph{{11th
  Inter-national Workshop on Algebraic and Combinatorial Coding Theory}}, 2008,
  pp. 192--198.

\bibitem{ratsaby2008estimate}
J.~Ratsaby, ``Estimate of the number of restricted integer-partitions,''
  \emph{Applicable Analysis and Discrete Mathematics}, pp. 222--233, 2008.

\bibitem{migler2004weight}
T.~Migler, K.~E. Morrison, and M.~Ogle, ``Weight and rank of matrices over
  finite fields,'' \emph{arXiv preprint math/0403314}, 2004.

\end{thebibliography}

\end{document}